\documentclass[hyper]{JHEP} 

\usepackage{epsfig}

\newcommand\fverb{\setbox\pippobox=\hbox\bgroup\verb}

\newcommand\fverbdo{\egroup\medskip\noindent%

            \fbox{\unhbox\pippobox}\ }

\newcommand\fverbit{\egroup\item[\fbox{\unhbox\pippobox}]}

\newbox\pippobox

\title{Non-Linear Massive Gravity with Additional
Primary Constraint and Absence of Ghosts}
\author{J. Kluso\v{n}\\
Department of
Theoretical Physics and Astrophysics\\
Faculty of Science, Masaryk University\\
Kotl\'{a}\v{r}sk\'{a} 2, 611 37, Brno\\
Czech Republic\\
E-mail: \email{klu@physics.muni.cz}}
\preprint{}

 \abstract{We complete the Hamiltonian analysis
 of  specific model of
 non-linear massive gravity that was
started in arXiv:1112.5267. We identify
the primary constraint  and
corresponding secondary  constraint. We
show that they are the second class
constraints and hence they lead to the
elimination of the additional scalar
mode. We also find that the remaining
constraints are the first class
constraints with the structure that
corresponds to the manifestly
diffeomorphism invariant theory.
Finally we determine the number of
physical degrees of freedom and we show
that it corresponds to the number of
physical modes of massive gravity. }

\def\bA{\mathbf{A}}
\def\bB{\mathbf{B}}

\def\mC{\mathcal{C}}
\def\be{\begin{equation}}

\def\ee{\end{equation}}

\def\bea{\begin{eqnarray}}
\def\tPi{\tilde{\Pi}}
\def\eea{\end{eqnarray}}

\def\tr{\mathrm{tr}\, }

\def\mH{\mathcal{H}}

\def\bz{\mathbf{z}}

\def\tr{\mathrm{Tr}}

\def\bx{\mathbf{x}}
\def\by{\mathbf{y}}

\newcommand{\mG}{\mathcal{G}}

\def \bA{\mathbf{A}}

\newcommand{\bT}{\mathbf{T}}

\newcommand{\mL}{\mathcal{L}}

\def\pb #1{\left\{#1\right\}}

\begin{document}
\section{Introduction and Summary}\label{first}
Recently  de Rham and Gobadadze
proposed in \cite{deRham:2010ik}  an
interesting formulation of the massive
gravity which is ghost free in the
decoupling limit. Then it was shown in
\cite{deRham:2010kj} that this action
that was written in the perturbative
form can be resumed into fully
non-linear action \footnote{For related
works, see
\cite{Sbisa':2012zk,Hinterbichler:2012cn,Paulos:2012xe,
Mirbabayi:2011xg,Sjors:2011iv,Burrage:2011cr,Gumrukcuoglu:2011zh,Berezhiani:2011mt,
Comelli:2011zm,Kluson:2011aq,Comelli:2011wq,Mohseni:2011vv,Gumrukcuoglu:2011ew,Hassan:2011zd,
Hassan:2011tf,D'Amico:2011jj,deRham:2011qq,deRham:2011pt,Gruzinov:2011mm,Koyama:2011yg,
Nieuwenhuizen:2011sq,
Koyama:2011xz,Chamseddine:2011bu,Volkov:2011an}.}.
The general analysis of the constraints
of given theory has been performed
 in \cite{Hassan:2011hr}. It was argued
 there that it is possible to perform
 such a redefinition of the shift
 function so that the
resulting theory still contains the
Hamiltonian constraint. Then it was
argued that the presence of this
constraint allows to eliminate the
scalar mode and hence the resulting
theory is the ghost free massive
gravity. However this analysis was
questioned in \cite{Kluson:2011qe}
where it was argued that it is possible
that this constraint is the second
class constraint so that the phase
space of given theory would be odd
dimensional. On the other hand in the
 paper \cite{Hassan:2011ea} very
nice analysis of the Hamiltonian
formulation of the most general gauge
fixed non-linear massive gravity
actions was performed  with an
important conclusion that the
Hamiltonian constraints has zero
Poisson brackets. Then the requirement
of the preservation of this constraint
during the time evolution of the system
implies an additional constraint. As a
result given theory has the right
number of constraints for the
construction of non-linear massive
gravity without additional scalar mode
\footnote{Alternative arguments for the
existence of an additional constraints
were given in \cite{Golovnev:2011nz}
even if the Hamiltonian analysis was
not complete and the minimal non-linear
massive gravity action was considered
only.}.

The Hamiltonian analysis of the
manifestly diffeomorphism invariant
non-linear massive gravity with
St\"{u}ckelberg fields was performed in
\cite{Kluson:2011rt} where
corresponding Hamiltonian was found.
Then using the observation firstly
published in \cite{Hassan:2012qv} it
was shown that this theory possesses
one primary constraint. Unfortunately
the presence of this constraint makes
the calculation of the Poisson brackets
between constraints very difficult due
to the absence of the inverse of the
matrix
$V^{AB}=g^{ij}\partial_i\phi^A\partial_j\phi^B$
and we were not able to perform this
analysis for the case of four
dimensional non-linear massive gravity.
On the other hand such an analysis was
performed for the case of two
dimensional non-linear massive gravity
with conclusions that there are no
physical degrees of freedom left with
agreement with
\cite{deRham:2011rn,Kluson:2011aq}.

In \cite{Kluson:2012gz} we analyzed the
model of the non-linear massive gravity
action introduced in
\cite{Golovnev:2011nz} written in the
St\"{u}ckelberg formalism.
 This
analysis was then reconsidered in
\cite{Hassan:2012qv} with conclusion
that this  theory is free from the
ghosts.

The goal of this paper is to complete
the analysis of the non-linear massive
gravity action presented in
\cite{Kluson:2011rt}. We find the
Hamiltonian for given theory and
identify primary constraints. Then we
rewrite the Hamiltonian to such a form
where the scalar part of the
Hamiltonian constraint will be
proportional to the trace of the square
root of the regular matrix. Then it
would be possible to use the standard
formula for the variation of the trace
of the square root of the regular
matrix and calculate corresponding
Poisson brackets. Then we can analyze
the requirement of the preservation of
the primary constraints during the time
evolution of the system and hence
identify corresponding  secondary
constraints. Finally we will check the
stability of all constraints during the
time evolution of the system. We find
that the Hamiltonian and diffeomorphism
constraints are still the first class
constraints and obey the basic
principles of geometrodynamics
\cite{Kuchar:1974es,Isham:1984sb,Isham:1984rz,Hojman:1976vp}.
On the other hand we show that the
additional  primary constraint together
with corresponding secondary constraint
are the second class constraints and
that these constrains could eliminate
one additional degree of freedom so
that the number of physical degrees of
freedom correspond to the case of the
massive gravity.
 In other words our results
are in full agreement with the
conclusion presented in
\cite{Hassan:2012qv}. However we  mean
that  result derived in this paper is
non-trivial and should be considered as
an independent check of the absence of
the ghosts in given theory due to the
fact that we analyze theory without
additional auxiliary fields so that the
Hamiltonian analysis presented her  is
different from the analysis presented
in \cite{Hassan:2012qv}.

We should also stress that  our
treatment  has its own limitation due
to the fact  that we restrict to the
case of one specific model of
non-linear massive gravity action. It
turns out that the extension of given
analysis to the more general form of
the non-linear massive gravity actions
is very difficult due to the
 complicated relation between
canonical momenta and time derivatives
of the scalar fields. Unfortunately we
were not able to find an inverse
mapping that would allow us to write
the Hamiltonian as a function canonical
variables in these cases. It  would be
very interesting to find such
Hamiltonian formulation and
corresponding primary constraints
between St\"{u}ckelberg fields even for
the most general form of the non-linear
massive gravity action. We hope to
return to this problem in future.

The structure of given note is as
follows. In the next section
(\ref{second}) we review some basic
facts about  the non-linear massive
gravity action in the formulation
presented in \cite{Kluson:2011rt}. Then
in section (\ref{third}) which is the
main body of this paper we perform
corresponding Hamiltonian analysis. We
also identify primary and the secondary
constraints and determine the number of
the physical degrees of freedom.

\section{Non-Linear Massive Gravity}\label{second}
 Our goal is to study non-linear
massive gravity action in the following
form \footnote{ We use notations
introduced in the paper
\cite{Kluson:2011rt}}
\begin{equation}\label{Smassive1}
S=M_p^2\int d^3\bx dt  N
\sqrt{g}\left[K_{ij}\mG^{ijkl}K_{kl}+{}^{(3)}R
- m^2 \tr_A \sqrt{\bA}\right] \ ,
\end{equation}
where we used   $3+1$ notation
\cite{Arnowitt:1962hi} \footnote{For
review, see \cite{Gourgoulhon:2007ue}.}
and write the four dimensional metric
components as
\begin{eqnarray}
\hat{g}_{00}=-N^2+N_i g^{ij}N_j \ ,
\quad \hat{g}_{0i}=N_i \ , \quad
\hat{g}_{ij}=g_{ij} \ ,
\nonumber \\
\hat{g}^{00}=-\frac{1}{N^2} \ , \quad
\hat{g}^{0i}=\frac{N^i}{N^2} \ , \quad
\hat{g}^{ij}=g^{ij}-\frac{N^i N^j}{N^2}
\ .
\nonumber \\
\end{eqnarray}
Note that  ${}^{(3)}R$ is
three-dimensional spatial curvature,
$K_{ij}$ is extrinsic curvature defined
as
\begin{equation}
K_{ij}=\frac{1}{2N} (\partial_t g_{ij}
-\nabla_i N_j-\nabla_j N_i) \ ,
\end{equation}
where $\nabla_i$ is covariant
derivative built from the metric
components $g_{ij}$ and $\mG^{ijkl}$ is
de Witt metric defined as
\begin{equation}
\mG^{ijkl}=\frac{1}{2}(g^{ik}g^{jl}+g^{il}g^{jk})-g^{ij}
g^{kl} \
\end{equation}
with inverse
\begin{equation}
\mG_{ijkl}=\frac{1}{2}
(g_{ik}g_{jl}+g_{il}g_{jk})-\frac{1}{2}g_{ij}g_{kl}
\ ,  \quad
\mG_{ijkl}\mG^{klmn}=\frac{1}{2}(\delta_i^m\delta_j^n+
\delta_i^n\delta_j^m) \ .
\end{equation}
Finally note that the matrix $\bA^A_{ \
B}$ is defined as
\begin{equation}
\bA^A_{ \ B}=
-\nabla_n\phi^A\nabla_n\phi_B+g^{ij}\partial_i\phi^A
\partial_j\phi_B \ ,
\quad \nabla_n\phi^A= \frac{1}{N}
(\partial_t\phi^A-N^i\partial_i \phi^A)
\
\end{equation}
and the trace defined in
(\ref{Smassive1}) is the trace over
Lorentz indices $A,B,C,\dots=0,1,2,3$.

We see that the action contains the
potential term which is the square root
of the matrix which can be defined as
\begin{equation}
(\sqrt{\bA})^A_{\ B}(\sqrt{\bA})^B_{ \
C}= \bA^A_{ \ C} \ .
\end{equation}
For further purposes it is crucial to
presume that $\bA$ is regular matrix.
Then when we perform the variation of
this expression and multiply by
$(\sqrt{\bA})^{-1}$ from the right we
obtain
\begin{equation}\label{deltabA}
\delta (\sqrt{\bA})^A_{ \ B}+
(\sqrt{\bA})^A_{ \ C}\delta
(\sqrt{\bA})^C_{ \ D}(
(\sqrt{\bA})^{-1})^D_{ \ B}=\delta
(\bA)^A_{ \ C} ((\sqrt{\bA})^{-1})^C_{
\ B} \ .
\end{equation}
Taking the trace the equation
(\ref{deltabA})  we immediately obtain
\footnote{Note also that due to the
matrix nature of objects $\bA$ and
$\bB$ the following relation is not
valid
\begin{equation}\label{sqrtAB}
\sqrt{\bA \bB}=\sqrt{\bA}\sqrt{\bB} \
\end{equation}
unless $\bA$ and $\bB$ commute. On the
other hand  since obviously $\bA$ and
$\bA^{-1}$ commute the equation
(\ref{sqrtAB}) gives
\begin{equation}
\sqrt{\bA}\sqrt{\bA^{-1}}=\mathbf{I} \
\end{equation}
which implies following important
relation
\begin{equation}
\left(\sqrt{\bA}\right)^{-1}=\sqrt{\bA^{-1}}
\ .
\end{equation}}
\begin{equation}\label{defdeltatrA}
\delta \tr_L\sqrt{\bA}= \frac{1}{2}
\delta \bA^A_{ \ B}
\left((\sqrt{\bA})^{-1}\right)^B_{ \ A}
\ .
\end{equation}
This is the key formula that is used in
the calculation of the Poisson brackets
as we will see in the next section.


\section{Hamiltonian
Analysis}\label{third}
 In this section
we perform the Hamiltonian analysis of
the action (\ref{Smassive1}). For the
General Relativity part of the action
the procedure is standard. Explicitly,
the momenta conjugate to $N,N^i$ are
the primary constraints of the theory
\begin{equation}
\pi_N(\bx)\approx 0 \ , \quad
\pi_i(\bx)\approx 0 \
\end{equation}
while the Hamiltonian takes the form
\begin{eqnarray}\label{HamGR}
H^{GR}&=&\int d^3\bx
(N\mH^{GR}_T+N^i\mH^{GR}_i) \ ,
\nonumber \\
\mH_T^{GR}&=&\frac{1}{\sqrt{g}M_p^2}
\pi^{ij}\mG_{ijkl}\pi^{kl}-M_p^2
\sqrt{g} {}^{(3)}R \ , \nonumber \\
\mH_i^{GR}&=& -2g_{ik}
\nabla_l\pi^{kl} \ , \nonumber \\
\end{eqnarray}
where $\pi^{ij}$ are momenta conjugate
to $g_{ij}$ with following non-zero
Poisson brackets
\begin{equation}
\pb{g_{ij}(\bx),\pi^{kl}(\by)}=
\frac{1}{2}\left(\delta_i^k\delta_j^l+\delta_i^l
\delta_j^k\right)\delta(\bx-\by) \ .
\end{equation}
Note also that $\pi_N,\pi_i$ have
following Poisson brackets with $N,N^i$
\begin{equation}
\pb{N(\bx),\pi_N(\by)}=\delta(\bx-\by)
\ , \quad \pb{N^i(\bx),\pi_j(\by)}=
\delta^i_j\delta(\bx-\by) \ .
\end{equation}
Now we proceed to the Hamiltonian
analysis of the scalar field part of
the action. Note that in $3+1$
formalism the matrix $\bA^A_{ \ B}$
takes the form
\begin{equation}
\bA^A_{\
B}=-\nabla_n\phi^A\nabla_n\phi_B+
g^{ij}\partial_i\phi^A\partial_j\phi_B
\equiv K^A_{ \ B}+V^A_{ \ B} \ ,
\end{equation}
where
\begin{eqnarray}
K^A_{ \
B}&=&-\nabla_n\phi^A\nabla_n\phi_B \ ,
\quad K_{AB}=\eta_{AC}K^C_{ \ B}=K_{BA}
\ , \nonumber \\
V^A_{ \
B}&=&g^{ij}\partial_i\phi^A\partial_j\phi_B
\ , \quad V^{AB}=V^A_{ \ C}\eta^{CB}=
V^{BA} \ . \nonumber \\
\end{eqnarray}
Then the  conjugate momenta $p_A$ are
equal to
\begin{eqnarray}\label{pAext}
p_A&=&-\frac{M_p^2
m^2}{2}\sqrt{g}\frac{\delta \bA^C_{ \
D}}{\delta
\partial_t\phi^A}(\bA^{-1/2})^D_{ \
C}=\nonumber
\\
&=&\frac{M_p^2m^2}{2}\sqrt{g}(\nabla_n\phi_C
(\bA^{-1/2})^C_{ \ A}+
\eta_{AC}(\bA^{-1/2})^C_{ \
B}\nabla_n\phi^B) \ , \quad
\bA^{-1/2}=(\sqrt{\bA})^{-1} \ .
\nonumber
\\
\end{eqnarray}
Note that using the symmetry of
$\bA_{AB}=\bA_{BA}$ we can write
(\ref{pAext})  in simpler form
\begin{equation}
p_A=M_p^2
m^2\sqrt{g}(\bA^{-1/2})_{AB}\nabla_n\phi^B
\ .
\end{equation}
Using this expression we derive
following relation
\begin{eqnarray}
\frac{1}{gM_p^4 m^4}p_A p_B&=&
(\bA^{-1/2})_{AC}(\nabla_n\phi^C\nabla_n\phi^D)
(\bA^{-1/2})_{DB}= \nonumber \\
&=& (\bA^{-1/2})_{AC}( V^{CD}-\bA^{CD})
(\bA^{-1/2})_{DB} \nonumber \\
\end{eqnarray}
which implies
\begin{equation}\label{PiAV}
\Pi_{AB}=(\bA^{-1/2})_{AC}V^{CD}(\bA^{-1/2})_{DB}
\ ,
\end{equation}
where we introduced the matrix
$\Pi_{AB}$ defined as
\begin{equation}
\Pi_{AB}=\frac{1}{gm^4 M_p^4}p_A
p_B+\eta_{AB} \ .
\end{equation}
Note that when we multiply (\ref{PiAV})
by $V$ from the right we obtain (we use
matrix notation)
\begin{equation}
\Pi V=(\bA^{-1/2}V)(\bA^{-1/2}V)
\end{equation}
which  implies
\begin{equation}\label{bAV1}
\bA^{-1/2}V= \sqrt{\Pi V} \ .
\end{equation}
This relation will be important below.
The crucial point  for the Hamiltonian
analysis of the non-linear massive
gravity is the fact that  $V^{AB}$ has
the rank $3$ as was firstly explicitly
stressed in \cite{Hassan:2012qv}. In
fact, if we introduce the $4\times 3$
matrix $W^A_{ \ i}=
\partial_i\phi^A$ and its transpose
matrix $(W^T)^i_{ \
A}=\partial_i\phi^A$ which is $3\times
4$ matrix we can write
\begin{equation}
V^{AB}=W^A_{ \ i}g^{ij}(W^T)_j^{ \ B} \ .
\end{equation}
Then since $W^A_{ \ i},g^{ij}$ have the
rank $3$ we obtain that $V^{AB}$ has
the rank $3$ as well. As a result $\det
V=0$. In other words $V$ is not
invertible matrix.

With the help of these results it is
easy to determine corresponding
Hamiltonian
\begin{eqnarray}\label{HamSC}
\mH^{sc}&=&\partial_t \phi^A
p_A-\mL_{sc}
=M_p^2m^2\sqrt{g}NV^{AB}(\bA^{-1/2})_{BA}+N^ip_A\partial_i\phi^A=
\nonumber \\
&=&NM_p^2m^2\sqrt{g}\tr_L\sqrt{\Pi V}
+N^i p_A\partial_i\phi^A\equiv
N\mH_T^{sc}+N^i\mH_i^{sc} \   \nonumber \\
\end{eqnarray}
using (\ref{bAV1}) and using an obvious
relation $\tr_L \sqrt{V\Pi}= \tr_L
\sqrt{\Pi V}$.  With the help of these
results we find the Hamiltonian for the
action (\ref{Smassive1}) in the form
\begin{equation}\label{Hor}
H=\int d^3\bx
(N\mH_T+N^i\mH_i+v^i\pi_i+v^N\pi_N+v_c\mC)
\ ,
\end{equation}
where
\begin{equation}
\mH_T=\mH_T^{GR}+\mH_T^{sc} \ , \quad
\mH_i=\mH_i^{GR}+\mH_i^{sc} \ ,
\end{equation}
and where $\pi_i\approx 0 \ ,
\pi_N\approx 0$ are the primary
constraints of the theory. Note also
that the Hamiltonian (\ref{Hor})
contains primary constraint $\mC$ whose
explicit form follows from (\ref{PiAV})
when we calculate the determinant of
the matrix $\Pi_{AB}$. Using
\begin{eqnarray}
& &\det \Pi_{AB}=
-\left(1+\frac{1}{gM_p^4m^4}p_A
p^A\right)
 \nonumber \\
\end{eqnarray}
and using (\ref{PiAV}) together with
the fact that $\det V=0$ we derive
\emph{primary constraint} $\mC$ in the
form
\begin{eqnarray}\label{detPi}
\mC:1+\frac{1}{gM_p^4m^4} p_A
p^A\approx 0 \ .
\nonumber \\
\end{eqnarray}
It is
also important to stress that using the
definition of $\Pi_{AB}$ and the
existence of the constraint $\mC$ we
obtain an important relation
\begin{equation}\label{pAPiAB}
p^A\Pi_{AB}= \left(\frac{1}{gM_p^4m^4}
p^Ap_A+1\right)p_B= \mC p_B\approx 0 \
.
\end{equation}
Now we analyze the requirement of the
preservation of the primary
constraints. As usual the requirement
of the preservation of the primary
constraints $\pi_N\approx 0 \ ,
\pi_i\approx 0$ implies an existence of
the secondary constraints
\begin{equation}\label{seccon}
\mH_T(\bx)\approx 0 \ , \quad
\mH_i(\bx)\approx 0 \ .
\end{equation}
For  further purposes we introduce  the
smeared form of these constraints
(\ref{seccon})
\begin{equation}\label{smearedconst}
\bT_T(N)=\int d^3\bx N\mH_T \ , \quad
\bT_S(N^i)=\int d^3\bx N^i\mH_i \ .
\end{equation}
It is not easy to determine the time
evolution of the constraint $\mC$ due
to the fact that $\Pi V$ is singular
matrix. To proceed let us express the
trace of the square root of the matrix
as power series in the form
\begin{equation}
\tr_L\sqrt{\Pi V}=\sum_n c_n \tr_L (\Pi
V)^n \ .
\end{equation}
Now we can write
\begin{eqnarray}\label{trLs}
\tr_L \Pi V&=&\Pi^{AB}\partial_i \phi^B
g^{ij}\partial_j\phi_A=
\partial_j\phi_A
\Pi^{AB}\partial_i\phi_Bg^{ij}\equiv
\tPi_j^{ \ j}\equiv \tr_s \tPi \ , \nonumber \\
\tr_L\Pi V\Pi V&=& (
\partial_i\phi_A\Pi^{AB}\partial_j\phi_B
g^{jk})(\partial_k \phi_C\Pi^{CD}
\partial_l\phi_D g^{li})=\tPi_i^{ \ k}
\tPi_k^{ \ i}\equiv \tr_s \tPi^2 \ ,
\nonumber \\
& &\vdots
 \nonumber \\
\end{eqnarray}
where the trace $\tr_s$ is the trace
over spatial indices
$i,j,k\dots=1,2,3$. Then with the help
of (\ref{trLs}) it is easy to see that
\begin{equation}
\tr_L\sqrt{\Pi V}= \tr_s \sqrt{\tPi} \
.
\end{equation}
Now $\tPi$ is $3\times 3$ matrix with
the rank equal to  $3$  which implies
an existence of the inverse matrix
$\tPi^{-1}$. As a result we can easily
determine the variation of the trace of
the square root of given matrix
\begin{equation}
\delta \tr_s\sqrt{\tPi}=
\frac{1}{2}\tr_s\delta \tPi
\sqrt{\tPi^{-1}} \ .
\end{equation}
Then we can determine following Poisson
brackets
\begin{eqnarray}\label{pbpphitPi}
& &\pb{p_A(\bx),\tr \sqrt{\tPi}(\by)}=
-\frac{\delta \tr \sqrt{\tPi}(\by)}{
\delta \phi^A(\bx)}= -\frac{1}{2}
\frac{\delta \tPi_i^{ \ j}(\by)}
{\delta
\phi^A(\bx)}\sqrt{\tPi^{-1}(\by)}_j^{ \
i}=\nonumber \\
&=&-\frac{1}{2}
(\partial_{y^i}\delta(\bx-\by)\eta_{AC}\Pi^{CD}
\partial_{y^k}\phi_Dg^{kj}+\partial_{y^i}
\phi_C\Pi^{CD}\eta_{DA}\partial_{y^k}
\delta(\bx-\by)g^{kj})(\by)
\sqrt{\tPi^{-1}(\by)}_j^{ \
i}\nonumber \\
&
&\pb{\phi^A(\bx),\tr\sqrt{\tPi}(\by)}=
\frac{\delta \sqrt{\tPi}(\by)}{\delta
p_A(\bx)}=\frac{1}{2} \frac{\delta
\tPi_i^{ \ j}(\by)}
{\delta p_A(\bx)}(\sqrt{\tPi^{-1}(\by)})^{ \ i}_{j}=\nonumber \\
&=&\frac{1}{2gm^4M_p^4} (\partial_i
\phi^A p_K\partial_k\phi^K+\partial_i
\phi^K p_K\partial_k
\phi^A)g^{kj}(\sqrt{\tPi^{-1}(\by)})^{
\ i}_{j}\delta(\bx-\by) \ . \nonumber
\\
\end{eqnarray}
Using these results we  find
\begin{eqnarray}\label{pbbTSmC}
\pb{\bT_S(N^i),\mC(\bx)}=-N^i\partial_i \mC(\bx) \   \nonumber \\
\end{eqnarray}
and also
\begin{eqnarray}
\pb{\bT_T(N),\mC(\bx)}=
-\frac{1}{gM_p^2m^2}
p^A\left(\partial_i[N\sqrt{g}\eta_{AC}\Pi^{CD}\partial_k\phi_D
g^{kj} \sqrt{\tPi^{-1}}^{ \
i}_{j}]\right.-
\nonumber \\
+\left.\partial_k[N\partial_i\phi_C\Pi^{CD}\eta_{DA}g^{kj}
\sqrt{\tPi^{-1}}^{ \
i}_{j}]\right)-\frac{2N}{M_p^6m^4g^{3/2}}p_Ap^A
g^{ij}\mG_{ijkl}\pi^{kl}=\nonumber
\frac{2N}{gM_p^2m^2}\mC^{II} \nonumber
\\
\end{eqnarray}
where we used (\ref{pAPiAB}) and where
we defined $\mC^{II}$ as
\begin{eqnarray}
\mC^{II}&=&
p_A\partial_i\Pi^{AB} \sqrt{g}
\partial_j\phi_B  \left(\sqrt{\tPi^{-1}}\right)^{ji}-\frac{2}{M_p^4m^2\sqrt{g}}p_Ap^A
g_{ij}\pi^{ji} \ ,
 \nonumber \\
 \end{eqnarray}
 where we defined
 $\left(\sqrt{\tPi^{-1}}\right)^{ij}=\left(\sqrt{\tPi^{-1}}\right)^{ji}=\sqrt{\tPi^{-1}}^i_{ \
 k}g^{kj}$.
 Now it is easy to see
 that the requirement
of the preservation of the constraint
$\mC$ during the time evolution of the
system implies following secondary
constraint
\begin{equation}
\partial_t
\mC=\pb{H,\mC}\approx
\pb{\bT_T(N),\mC}=\frac{N}{M_p^2m^2
g}\mC^{II}\approx 0 \ .
\end{equation}
In summary,  the theory possesses
following collection of the primary
constraints $\pi_N\approx 0 \ ,
\pi_i\approx 0 \ , \mC\approx 0$ and
secondary constraints $\mH_T\approx 0 \
, \mH_i\approx 0 \ , \mC^{II}\approx
0$. As a result the total Hamiltonian
has the form
\begin{equation}
H_T=\int d^3\bx (N\mH_T+N^i\mH_i+
v_N\pi_N+v^i\pi_i+v_\mC \mC+
\Gamma^i\mH_i+\Gamma_{\mC}\mC^{II}) \ ,
\end{equation}
where
$v_N,v^i,v_\mC,\Gamma^i,\Gamma_{\mC}$
are corresponding Lagrange multipliers.

As the final step we  have to analyze
the preservation of all constraints.
Note that in case of the General
Relativity part of the constraints we
have following Poisson brackets
\begin{eqnarray}\label{PBGRCON}
\pb{\mH_T^{GR}(\bx),\mH_T^{GR}(\by)}&=&
-\left[\mH^i_{GR}(\bx)\frac{\partial}{\partial
x^i}
\delta(\bx-\by)-\mH^i_{GR}(\by)\frac{\partial}{\partial
y^i} \delta(\bx-\by)\right] \ , \nonumber \\
\pb{\mH_T^{GR}(\bx),\mH_i^{GR}(\by)}&=&
\mH^{GR}_T(\by)\frac{\partial}{\partial
x^i}\delta(\bx-\by) \ , \nonumber \\
\pb{\mH^{GR}_i(\bx),\mH_j^{GR}(\by)}&=&
\left[\mH_j^{GR}(\bx)\frac{\partial}{\partial
x^i} \delta(\bx-\by)-\mH_i(\by)
\frac{\partial}{\partial y^j}
\delta(\bx-\by)\right] \ . \nonumber \\
\end{eqnarray}
The calculation of the Poisson brackets
that contains scalar phase space
degrees of freedom is more involved.
However it is easy to find the Poisson
bracket between generators of spatial
diffeomorphisms
\begin{equation}
\pb{\mH^{sc}_i(\bx),\mH_j^{sc}(\by)}=
\left[\mH_j^{sc}(\bx)\frac{\partial}{\partial
x^i} \delta(\bx-\by)-\mH^{sc}_i(\by)
\frac{\partial}{\partial y^j}
\delta(\bx-\by)\right] \
\end{equation}
that together with the  Poisson bracket
on the third line in  (\ref{PBGRCON})
implies following form of Poisson
bracket between smeared form of the
diffeomorphism constraints
\begin{equation}\label{pbbTSS}
\pb{\bT_S(N^i),\bT_S(M^j)}=
\bT_S(N^j\partial_j M^i-M^j\partial_j
N^i) \ .
\end{equation}
It is also easy to see that
\begin{equation}\label{pbbTST}
\pb{\bT_S(N^i),\bT_T(N)}=
\bT_T(\partial_k N N^k) \ .
\end{equation}
Now we proceed to the calculation of
the Poisson bracket
$\pb{\bT_T(N),\bT_T(M)}$. By definition
we have
\begin{eqnarray}\label{bTNM}
& &
\pb{\bT^{sc}_T(N),\bT^{sc}_T(M)}=\int
d^3\bx d^3\by N(\bx)
\pb{\mH_T^{sc}(\bx),\mH_T^{sc}(\by)}M(\by)=
\nonumber \\
&=&-M^4_p m^4\int d^3\bx d^3\by
 d^3\bz
 N(\bx)M(\by) \left(
\sqrt{g}(\bx) \frac{\delta (\sqrt{\tPi
})_{i}^{ \ i}(\bx)} {\delta p_X(\bz)}
\frac{\delta (\sqrt{\tPi})_j^{\
j}(\by)}{\delta
\phi^X(\bz)}\sqrt{g}(\bx) -
\right. \nonumber \\
& &\left. - \sqrt{g}(\by)\frac{\delta
(\sqrt{\tPi})_{j}^{ \ j}(\by)}
 {\delta p_X(\bz)}
\frac{\delta (\sqrt{\tPi})_i^{\
i}(\bx)}{\delta
\phi^X(\bz)}\sqrt{g}(\bx) \right)=
\nonumber \\
&=&\bT^{sc}_S((N\partial_j
M-M\partial_j)g^{ji}) \ ,  \nonumber \\
\end{eqnarray}
where we used  (\ref{pbpphitPi}). This
result together with (\ref{PBGRCON})
implies \footnote{It is important to
stress that
$\pb{\bT_T^{GR}(N),\bT_T^{sc}(M)}+
\pb{\bT_T^{sc}(M),\bT_T^{GR}(N)}=0$ due
to the fact that $\bT_T^{sc}$ does not
depend on the spatial derivatives of
$g_{ij}$.}
\begin{equation}
\pb{\bT_T(N),\bT_T(M)}=\bT_S(
(N\partial_j M-M\partial_j)g^{ji}) \ .
\end{equation}
Now we have to determine whether all
constraints  are preserved during the
time evolution of the system. Let us
now start with the primary constraints
$\pi^N,\pi_i,\mC\approx 0$. The case of
the constraints $\pi_N\approx 0 \ ,
\pi_i\approx 0$ is trivial.
Further, the requirement of the
preservation of the constraint $\mC$
gives
\begin{eqnarray}
\partial_t \mC(\bx)&=&\pb{H_T,\mC(\bx)}\approx
\int d^3\by\left(
\pb{N(\by)\mH_T(\by),\mC(\bx)}+
\Gamma_{\mC}(\by)\pb{\mC^{II}(\by),\mC(\bx)}\right)=
\nonumber
\\
&=&\frac{2N}{M_p^2m^2 g}\mC^{II}(\bx)+
\int d^3\by
\Gamma_{\mC}(\by)\pb{\mC^{II}(\by),\mC(\bx)}\approx
\nonumber \\
&\approx &  \int d^3\by
\Gamma_{\mC}(\by)\pb{\mC^{II}(\by),\mC(\bx)}=0
\nonumber \\
\end{eqnarray}
that has the solution $\Gamma_\mC=0$
using the fact that
\begin{eqnarray}\label{mCIImC}
\pb{\mC^{II}(\bx),\mC(\by)}= 2p_A(\bx)
\partial_i\Pi^{AB}(\bx) \sqrt{g}(\bx)
\left(\sqrt{\tPi^{-1}}\right)^{ji}(\bx)
\partial_j\delta(\bx-\by)
\frac{p_B(\by)}{M_p^4m^4 g(\by)}+\dots
\nonumber \\
\end{eqnarray}
does not vanish on the constraint
surface. Note that  $\dots$ mean
additional terms that arise from the
explicit calculations of given  Poisson
brackets. In other words $\mC$ and
$\mC^{II}$ are the second class
constraints.

 Now we come to the
requirement of the preservation of the
secondary constraints. Let us begin
with the diffeomorphism constrains
$\mH_i$ or their smeared forms. Since
$\mC^{II}\approx 0$ is manifestly
diffeomorphism invariant we have
$\pb{\bT_S(N^i),\mC^{II}}\approx 0$ and
also using  (\ref{pbbTSmC}) together
with   (\ref{pbbTSS}) and
(\ref{pbbTST})
 we find
that $\mH_i$ is preserved during the
time evolution of the system. In  case
of $\mH_T$ we find that its time
development is governed by the equation
\begin{eqnarray}
\partial_t\mH_T(\bx)&=&\pb{H_T,\mH_T(\bx)}
\approx \int d^3\by\left(
\pb{v_\mC\mC(\by),\mH_T(\bx)}+
\Gamma_\mC\pb{\mC^{II}(\by),\mH_T(\bx)}\right)=
\nonumber\\
&=&\int d^3\by
\pb{v_\mC\mC(\by),\mH_T(\bx)}=
\frac{2v_\mC(\bx)}{M_p^2m^2 g}
\mC^{II}(\bx)\approx 0 \nonumber \\
\end{eqnarray}
using (\ref{bTNM}) and also  using the
fact that $\Gamma_\mC=0$. In other
words $\mH_T$ is also preserved during
the time evolution of the system
without any restriction on the lapse
function $N$.
Finally the requirement of the
preservation of the constraint
$\mC^{II}$ has the form
\begin{eqnarray}
\partial_t
\mC^{II}(\bx)&=&\pb{H_T,\mC^{II}(\bx)}=
\nonumber \\
&=&\int
d^3\by\left(\pb{N\mH_T(\by),\mC^{II}(\bx)}+
v_\mC(\by)\pb{\mC(\by),\mC^{II}(\bx)}\right)=0
\nonumber \\
\end{eqnarray}
using the fact that $\Gamma_\mC=0$.
Then with the help of the equation
(\ref{mCIImC}) we can argue that this
solution can be solved for  $v_\mC$ at
least in principle.

Let us outline our results. We have
following first class constraints
$\pi_N\approx 0 \ , \pi_i\approx 0 \ ,
\mH_i\approx 0 \ , \mH_T\approx 0$
together with the second class
constraints $\mC\approx 0 \ ,
\mC^{II}\approx 0$. Then we also
 have $10$ metric components
 $g_{ij},N,N^i$
and corresponding conjugate momenta
$\pi^{ij},\pi_N,\pi_i$
 and
$4$ scalar fields $\phi^A$  with
conjugate momenta $p_A$. In general we
have $D=28$ phase space degrees of
freedom. On the other hand we have
$S=2$  second class constraints and
$F=8$ first class constraints. As a
result the number of physical degrees
of freedom is equal to
\cite{Henneaux:1992ig}
\begin{equation}
N_{p.d}=(D-S-2F)=10 \
\end{equation}
which is the correct number of the
physical degrees of freedom of the
massive gravity. In other words $\mC$
and $\mC^{II}$ eliminate the ghost
field and corresponding conjugate
momenta at least in principle.

 \noindent {\bf
Acknowledgements:} I would like to
thank to F. Hassan for very useful
discussion.
 This work   was
supported by the Grant agency of the
Czech republic under the grant
P201/12/G028. \vskip 5mm

\end{document}